\documentclass[12pt]{article}
\usepackage{amsmath}
\usepackage{amssymb}

\numberwithin{equation}{section}
\newcommand{\chh}{{\cal H}}
\newcommand{\crr}{{\cal R}}
\newcommand{\cii}{{\cal I}}
\newcommand{\css}{{\cal S}}
\newcommand{\cxx}{{\cal X}}

\title{
EXCHANGE OPERATOR FORMALISM FOR AN INFINITE FAMILY OF SOLVABLE AND INTEGRABLE QUANTUM SYSTEMS ON A PLANE}
\author{C. QUESNE\\
{\small \sl Physique Nucl\'eaire Th\'eorique et Physique Math\'ematique,}
\\ {\small \sl Universit\'e Libre de Bruxelles, Campus de la Plaine CP229,} \\ 
{\small \sl Boulevard~du Triomphe, B-1050 Brussels, Belgium} \\
{\small \sl cquesne@ulb.ac.be}}
\date{ }
\begin{document}
\baselineskip=22pt plus 1pt minus 1pt
\maketitle

\begin{abstract} 
The exchange operator formalism in polar coordinates, previously considered for the Calogero-Marchioro-Wolfes problem, is generalized to a recently introduced, infinite family of exactly solvable and integrable Hamiltonians $H_k$, $k=1$, 2, 3,~\ldots, on a plane. The elements of the dihedral group $D_{2k}$ are realized as operators on this plane and used to define some differential-difference operators $D_r$ and $D_{\varphi}$. The latter serve to construct $D_{2k}$-extended and invariant Hamiltonians $\chh_k$, from which the starting Hamiltonians $H_k$ can be retrieved by projection in the $D_{2k}$ identity representation space.   
\end{abstract}

\noindent
Running head: Exchange Operator Formalism


%
%
\newpage
\section{Introduction}

In a recent work, an infinite family of exactly solvable and integrable quantum Hamiltonians $H_k$, $k=1$, 2, 3,~\ldots, on a plane has been introduced \cite{tremblay09a}. Such a family includes all previously known Hamiltonians with the above properties, containing rational potentials and allowing separation of variables in polar coordinates. These correspond to the Smorodinsky-Winternitz (SW) system ($k=1$) \cite{fris, winternitz}, the rational $BC_2$ model ($k=2$) \cite{olsha81, olsha83}, and the Calogero-Marchioro-Wolfes (CMW) model ($k=3$) \cite{calogero74, wolfes} (reducing in a special case to the three-particle Calogero one \cite{calogero69}). Furthermore, it has been conjectured (and proved for the first few cases) that all members of the family are also superintegrable. In agreement with such a conjecture, all bounded classical trajectories have been shown to be closed and the classical motion to be periodic \cite{tremblay09b}.\par
%
%
Since the pioneering work of Olshanetsky and Perelomov \cite{olsha81, olsha83} on the integrability of Calogero-Sutherland type $N$-body models, {\em i.e.}, the existence of $N$ well-defined, commuting integrals of motion including the Hamiltonian, there have been several studies of such a problem using various approaches (see, {\em e.g.}, Refs.\ \cite{bordner, oshima} for some recent ones). One of the most interesting methods is based on the use of some differential-difference operators or covariant derivatives, known in the mathematical literature as Dunkl operators \cite{dunkl}. These operators were independently rediscovered by Polychronakos \cite{poly} and Brink {\em et al.} \cite{brink} in the context of the $N$-body Calogero model. Later on, they were generalized to the CMW model \cite{cq} and, in such a context, an interesting exchange operator formalism in polar coordinates was introduced \cite{khare}.\par
%
%
Since the CMW Hamiltonian is one of the members of the infinite family of Hamiltonians $H_k$, $k=1$, 2, 3,~\ldots, considered in Ref.\ \cite{tremblay09a}, it is worthwhile to extend the latter formalism to the whole family and to study some of its consequences. This is the purpose of this letter. To solve the problem, we shall have to distinguish between odd and even $k$ values and to prove several nontrivial trigonometric identities.\par
%
%
\section{\boldmath Odd $k$ Hamiltonians}

Let us consider the subfamily of Hamiltonians
\begin{equation}
  H_k = - \partial_r^2 - \frac{1}{r} \partial_r - \frac{1}{r^2} \partial_{\varphi}^2 + \omega^2 r^2 +
  \frac{k^2}{r^2} [a(a-1) \sec^2 k\varphi + b(b-1) \csc^2 k\varphi],  \label{eq:H-k}
\end{equation}
corresponding to $k=1$, 3, 5,~\ldots. Here $\omega$, $a$, $b$ are three parameters such that $\omega > 0$, $a(a-1) > - 1/(4k^2)$, $b(b-1) > - 1/(4k^2)$, and the configuration space is given by the sector $0 \le r < \infty$, $0 \le \varphi \le \pi/(2k)$. In cartesian coordinates  $x = r \cos \varphi$, $y = r \sin \varphi$, the Hamiltonian (\ref{eq:H-k}) can be rewritten as
\begin{align}
  H_1 & = - \partial_x^2 - \partial_y^2 + \omega^2 (x^2 + y^2) + \frac{a(a-1)}{x^2} + \frac{b(b-1}{y^2}, \\
  H_3 & = - \partial_x^2 - \partial_y^2 + \omega^2 (x^2 + y^2) + 9 (x^2 + y^2)^2 \left(\frac{a(a-1}{x^2
        (3y^2 - x^2)^2} + \frac{b(b-1)}{y^2 (3x^2 - y^2)^2}\right),  \label{eq:H-3} 
\end{align}  
and more and more complicated expressions as $k$ is increasing. $H_1$ is known as the SW Hamiltonian \cite{fris, winternitz}, while $H_3$ is the relative motion Hamiltonian in the CMW problem, as shown below.\par
%
%
\subsection{\boldmath The $k=3$ case}

The three-particle Hamiltonian of the CMW problem is given by \cite{wolfes}
\begin{equation}
\begin{split}
  H_{\rm CMW} &= \sum_{i=1}^3 (- \partial_i^2 + \omega^2 x_i^2) + 2a(a-1) \left(\frac{1}{x_{12}^2} + 
      \frac{1}{x_{23}^2} + \frac{1}{x_{31}^2}\right) \\
  & \quad + 6b(b-1) \left(\frac{1}{y_{12}^2} + \frac{1}{y_{23}^2}
      + \frac{1}{y_{31}^2}\right),
\end{split}  \label{eq:H-CMW}
\end{equation}
where $x_i$, $i=1$, 2, 3, denote the particle coordinates, $x_{ij} = x_i - x_j$, $i \ne j$, and $y_{ij} = x_i + x_j - 2 x_k$, $i \ne j \ne k \ne i$. The range of the particle coordinates is appropriately restricted as explained in Ref.\ \cite{cq}. In terms of the variables $x = x_{12}/\sqrt{2}$, $y = y_{12}/\sqrt{6}$, and $X = (x_1 + x_2 + x_3)/\sqrt{3}$, the Hamiltonian can be separated into a centre-of-mass Hamiltonian $H_{\rm cm} = - \partial_X^2 + \omega^2 X^2$ and a relative one $H_{\rm rel}$, coinciding with $H_3$ given in (\ref{eq:H-3}).\par
%
%
The CMW Hamiltonian (\ref{eq:H-CMW}) is known \cite{olsha81, olsha83} to be related to the $G_2$ Lie algebra, whose Weyl group is the dihedral group $D_6$. The 12 operators of the latter can be realized either in terms of the particle permutation operators $K_{ij}$ and the inversion operator $I_r$ in relative coordinate space \cite{cq} or in terms of the rotation operator $\crr = \exp\left(\frac{1}{3} \pi \partial_{\varphi}\right)$ through angle $\pi/3$ in the plane $(r, \varphi)$ and the operator $\cii = \exp({\rm i} \pi \varphi \partial_{\varphi})$ changing $\varphi$ into $- \varphi$ \cite{khare}. These exchange operators can then be used to extend the partial derivatives $\partial_i$ or $\partial_r$, $\partial_{\varphi}$ into differential-difference operators $D_i$ or $D_r$, $D_{\varphi}$, respectively.\par
%
%
In terms of the former
\begin{equation}
  D_i = \partial_i - a \sum_{j \ne i} \frac{1}{x_{ij}} K_{ij} - b \Biggl(\sum_{j\ne i} \frac{1}{y_{ij}} K_{ij} -
  \sum_{\substack{j,k \\ i \ne j \ne k \ne i}} \frac{1}{y_{jk}} K_{jk}\Biggr) I_r,   
\end{equation}
the latter can be defined as
\begin{equation}
  D_r = - \sqrt{\frac{2}{3}} \sum_i \sin \left(\varphi - i \frac{2\pi}{3}\right) D_i, \qquad 
  D_{\varphi} = - \sqrt{\frac{2}{3}}\, r \sum_i \cos \left(\varphi - i \frac{2\pi}{3}\right) D_i.  \label{eq:D-D}
\end{equation}
On using the correspondences $K_{ij} \leftrightarrow \crr^{2k+3} \cii$ and $K_{ij} I_r \leftrightarrow \crr^{2k} \cii$ with $(ijk) = (123)$, Eq.\ (\ref{eq:D-D}) can be rewritten as
\begin{equation}
  D_r = \partial_r - \frac{1}{r} (a\crr + b) (1 + \crr^2 + \crr^4) \cii,
\end{equation}
\begin{equation}
\begin{split}
  D_{\varphi} &= \partial_{\varphi} + a \left[\tan \varphi\, \crr^3 + \tan\left(\varphi + \frac{\pi}{3}\right) 
       \crr^5 + \tan\left(\varphi + \frac{2\pi}{3}\right) \crr\right] \cii \\
  & \quad - b \left[\cot \varphi + \cot\left(\varphi + \frac{\pi}{3}\right) \crr^2
       + \cot\left(\varphi + \frac{2\pi}{3}\right) \crr^4\right] \cii.
\end{split}  \label{eq:D-phi}
\end{equation}
Note that the small discrepancies existing between Eqs.\ (\ref{eq:D-D}) -- (\ref{eq:D-phi}) and the corresponding expressions (13) and (15) of Ref.\ \cite{khare} come from the fact that here we use the conventional definition of polar coordinates, while our former work was based on Wolfes' definition \cite{wolfes}.\par
%
%
We can now build on this exchange operator formalism in polar coordinates to go further and extend the Hamiltonian (\ref{eq:H-CMW}) itself. As a first step, we note that from the characteristic relations of $D_6$,
\begin{equation}
  \crr^6 = \cii^2 = 1, \qquad \cii \crr = \crr^5 \cii, \qquad \crr^{\dagger} = \crr^5, \qquad \cii^{\dagger} =
  \cii,  \label{eq:prop-1}
\end{equation}
it is easy to prove that $D_r$ and $D_{\varphi}$ satisfy the equations
\begin{equation}
  D_r^{\dagger} = - D_r - \frac{1}{r} [1 + 2(a\crr + b) (1 + \crr^2 + \crr^4) \cii], \qquad \crr D_r = D_r \crr, 
  \qquad \cii D_r = D_r \cii,  \label{eq:prop-2}
\end{equation}
\begin{equation}
  D_{\varphi}^{\dagger} = - D_{\varphi}, \qquad \crr D_{\varphi} = D_{\varphi} \crr, \qquad \cii D_{\varphi} =
  - D_{\varphi} \cii.  \label{eq:prop-3}
\end{equation}
It may be observed that the first relation in (\ref{eq:prop-3}) is similar to that fulfilled by the partial derivative operator $\partial_{\varphi}$. However, the first relation in (\ref{eq:prop-2}) differs from the corresponding result for $\partial_r$, namely $\partial_r^{\dagger} = - \partial_r - \frac{1}{r}$. Furthermore, in contrast with $\partial_r$ and $\partial_{\varphi}$ which commute with one another, $D_r$ and $D_{\varphi}$ satisfy the commutation relation
\begin{equation}
  [D_r, D_{\varphi}] = - \frac{2}{r} (a\crr + b) (1 + \crr^2 + \crr^4) \cii D_{\varphi}.  \label{eq:prop-4}
\end{equation}
\par
%
%
The next stage consists in expressing $D_{\varphi}^2$ in terms of $\varphi$, $\partial_{\varphi}$, $\crr$, and $\cii$. This can be done using Eqs.\ (\ref{eq:prop-1}) and (\ref{eq:prop-3}), as well as some well-known trigonometric identities. The result reads
\begin{equation}
\begin{split}
  D_{\varphi}^2 &= \partial_{\varphi}^2 - \Bigl[\sec^2 \varphi\, a(a - \crr^3 \cii) + \sec^2 \Bigl(\varphi + 
       \frac{\pi}{3}\Bigr) a(a - \crr^5 \cii) \\
  & \quad + \sec^2 \Bigl(\varphi + \frac{2\pi}{3}\Bigr) a(a - \crr \cii)\Bigr] - \Bigl[\csc^2 \varphi\, b(b - \cii)\\ 
  & \quad + \csc^2 \Bigl(\varphi + \frac{\pi}{3}\Bigr) b(b - \crr^2 \cii) +
       \csc^2 \Bigl(\varphi + \frac{2\pi}{3}\Bigr) b(b - \crr^4 \cii)\Bigr] \\
  & \quad + 3 (a^2 + b^2 + 2ab \crr) (1 + \crr^2 + \crr^4).  
\end{split}  \label{eq:prop-5}
\end{equation}
\par
%
%
{}Finally, we may introduce some generalized CMW Hamiltonian, defined by
\begin{equation}
  \chh_{\rm CMW} = H_{\rm cm} + \chh_{\rm rel},
\end{equation}
where
\begin{equation}
\begin{split}
  \chh_{\rm rel} &= \chh_3 = - \partial_r^2 - \frac{1}{r} \partial_r - \frac{1}{r^2} [D_{\varphi}^2 - 3 (a^2
      + b^2 + 2ab \crr) (1 + \crr^2 + \crr^4)] + \omega^2 r^2 \\
  &= - D_r^2 - \frac{1}{r} [1 + 2 (a\crr + b) (1 + \crr^2 + \crr^4) \cii] D_r - \frac{1}{r^2} D_{\varphi}^2 +
      \omega^2 r^2.
\end{split}
\end{equation}
Such a $D_6$-extended Hamiltonian is endowed with two interesting properties: {\em (i)} it is left invariant under $D_6$ and {\em (ii)} its projection in the representation space of the $D_6$ identity representation, obtained by replacing both $\crr$ and $\cii$ by 1, gives back the starting CMW Hamiltonian.\par
%
%
\subsection{\boldmath Generalization to other odd $k$ values}

We now plan to show that the formalism developed for $k=3$ in Sec.\ 2.1 can be extended to any other odd $k$ value (including $k=1$). For such a purpose, let us introduce the two operators $\crr = \exp\left(\frac{1}{k} \pi \partial_{\varphi}\right)$ and $\cii = \exp({\rm i} \pi \varphi \partial_{\varphi})$, satisfying the defining relations
\begin{equation}
  \crr^{2k} = \cii^2 = 1, \qquad \cii \crr = \crr^{2k-1} \cii, \qquad \crr^{\dagger} = \crr^{2k-1}, \qquad 
  \cii^{\dagger} = \cii  \label{eq:prop-1-bis}  
\end{equation}
of the dihedral group $D_{2k}$, whose elements may be realized as $\crr^i$ and $\crr^i \cii$, $i=0$, 1, \ldots, $2k-1$. It is then straightforward to show that the differential-difference operators
\begin{align}
  D_r &= \partial_r - \frac{1}{r} (a\crr + b) \left(\sum_{i=0}^{k-1} \crr^{2i}\right) \cii,  \label{eq:D-r-bis}  \\
  D_{\varphi} &= \partial_{\varphi} + a \sum_{i=0}^{k-1} \tan\left(\varphi + i \frac{\pi}{k}\right) \crr^{k+2i} \cii
        - b \sum_{i=0}^{k-1} \cot\left(\varphi + i \frac{\pi}{k}\right) \crr^{2i} \cii \label{eq:D-phi-bis}
\end{align}
still fulfil Eq.\ (\ref{eq:prop-3}), while Eqs .\ (\ref{eq:prop-2}) and (\ref{eq:prop-4}) are generalized into
\begin{equation}
  D_r^{\dagger} = - D_r - \frac{1}{r} \left[1 + 2(a\crr + b) \left(\sum_{i=0}^{k-1} \crr^{2i} \right) \cii\right], 
  \qquad \crr D_r = D_r \crr, \qquad \cii D_r = D_r \cii  \label{eq:prop-2-bis}
\end{equation}
and
\begin{equation}
  [D_r, D_{\varphi}] = - \frac{2}{r} (a\crr + b) \left(\sum_{i=0}^{k-1} \crr^{2i} \right) \cii D_{\varphi}, 
  \label{eq:prop-4-bis} 
\end{equation}
respectively.\par
%
%
The extension of Eq.\ (\ref{eq:prop-5}), however, turns out to be more tricky, because in the calculation there appear some involved sums of trigonometric functions. The simplest ones,
\begin{equation}
  \sum_{i=0}^{k-1} \sec^2 \left(\varphi + i \frac{\pi}{k}\right) = k^2 \sec^2 k\varphi, \qquad 
  \sum_{i=0}^{k-1} \csc^2 \left(\varphi + i \frac{\pi}{k}\right) = k^2 \csc^2 k\varphi,  \label{eq:identity-1}
\end{equation}
have been proved (under a slightly different form) in Ref.\ \cite{jakubsky} by using some elegant method. Inspired by this type of approach, we have demonstrated the three additional identities
\begin{equation}
  \sum_{i=0}^{k-1} \tan \left[\varphi + (i+j) \frac{\pi}{k}\right] \tan \left[\varphi + (i+2j) \frac{\pi}{k}\right]
       = - k, \qquad j = 1, 2, \ldots, k-1, 
\end{equation}
\begin{equation}
  \sum_{i=0}^{k-1} \cot \left[\varphi + (i+j) \frac{\pi}{k}\right] \cot \left[\varphi + (i+2j) \frac{\pi}{k}\right]
       = - k, \qquad j = 1, 2, \ldots, k-1, 
\end{equation}
\begin{equation}
\begin{split}
  & \sum_{i=0}^{k-1} \Bigl\{\tan \Bigl[\varphi + (i+j) \frac{\pi}{k}\Bigr] \cot \Bigl[\varphi + (i+2j) \frac{\pi}{k}
       \Bigr]  \\
  & \quad+ \cot \Bigl[\varphi + (i+j) \frac{\pi}{k}\Bigr] \tan \Bigl[\varphi + (i+2j) \frac{\pi}{k}\Bigr]\Bigr\} 
  = 2k, \qquad j = 1, 2, \ldots, k-1.
\end{split}  \label{eq:identity-4}
\end{equation}
It is worth stressing that in Eqs.\ (\ref{eq:identity-1}) -- (\ref{eq:identity-4}), $k$ is restricted to odd values. As it has been shown in Ref.\ \cite{jakubsky}, the counterpart of the first relation in Eq.\ (\ref{eq:identity-1}), for instance, looks entirely different for even $k$.\par
%
%
On taking advantage of such results, we arrive at the equation
\begin{equation}
\begin{split}
  D_{\varphi}^2 &= \partial_{\varphi}^2 - \sum_{i=0}^{k-1} \sec^2 \left(\varphi + i \frac{\pi}{k}\right) a (a -
       \crr^{k+2i} \cii) - \sum_{i=0}^{k-1} \csc^2 \left(\varphi + i \frac{\pi}{k}\right) b (b - \crr^{2i} \cii) \\
  & \quad + k (a^2 + b^2 + 2ab \crr) \sum_{i=0}^{k-1} \crr^{2i},   
\end{split}
\end{equation}
from which we can build a $D_{2k}$-extended Hamiltonian
\begin{equation}
\begin{split}
  \chh_k &= - \partial_r^2 - \frac{1}{r} \partial_r - \frac{1}{r^2} \left[D_{\varphi}^2 - k (a^2 + b^2 + 2ab \crr)
        \sum_{i=0}^{k-1} \crr^{2i}\right] + \omega^2 r^2 \\
  &= - D_r^2 - \frac{1}{r} \left[1 + 2 (a\crr + b) \left(\sum_{i=0}^{k-1} \crr^{2i}\right) \cii\right] D_r
        - \frac{1}{r^2} D_{\varphi}^2 + \omega^2 r^2, 
\end{split}  \label{eq:gen-H-k}
\end{equation}
left invariant under $D_{2k}$ and giving back $H_k$ by projection in the identity representation space.\par
%
%
\section{\boldmath Even $k$ Hamiltonians}

Considering next the subfamily of Hamiltonians (\ref{eq:H-k}) for $k=2$, 4, 6,~\ldots, we note that the only member known in the literature before the work of Ref.\ \cite{tremblay09a} was the $BC_2$ Hamiltonian $H_2$, whose expression in cartesian coordinates reads \cite{olsha83}
\begin{equation}
  H_2 = - \partial_x^2 - \partial_y^2 + \omega^2 (x^2 + y^2) + 2a(a-1) \left(\frac{1}{(x-y)^2} + 
  \frac{1}{(x+y)^2}\right) + b(b-1) \left(\frac{1}{x^2} + \frac{1}{y^2}\right).  \label{eq:H-2} 
\end{equation}
We shall therefore start our study of the even $k$ case by reviewing this example.\par
%
%
\subsection{\boldmath The $k=2$ case}

{}For the $BC_2$ model, the Weyl group is the dihedral group $D_4$, whose eight operators can be realized either in terms of the operator $K$ interchanging $x$ with $y$ and the reflection operators $I_x: x \to - x$, $I_y: y \to -y$, or in terms of the rotation operator $\crr = \exp\left(\frac{\pi}{2} \partial_{\varphi}\right)$ through angle $\pi/2$ and the operator $\cii = \exp({\rm i} \pi \varphi \partial_{\varphi})$ changing $\varphi$ into $- \varphi$.\par
%
%
{}From the former, we can construct differential-difference operators in cartesian coordinates
\begin{align}
  D_x &= \partial_x - a \left(\frac{1}{x-y} + \frac{1}{x+y} I_x I_y\right) K - \frac{b}{x} I_x,  \\
  D_y &= \partial_y + a \left(\frac{1}{x-y} - \frac{1}{x+y} I_x I_y\right) K - \frac{b}{y} I_y.
\end{align}
By proceeding as in the CMW model \cite{khare}, we can then introduce corresponding operators in polar coordinates
\begin{equation}
  D_r = \cos \varphi D_x + \sin \varphi D_y, \qquad D_{\varphi} = r (- \sin \varphi D_x + \cos \varphi D_y). 
\end{equation}
On taking advantage of the correspondences $K \leftrightarrow \crr^3 \cii$, $I_x \leftrightarrow \crr^2 \cii$, $I_y \leftrightarrow \cii$, $I_x I_y K \leftrightarrow \crr \cii$, such new operators can be rewritten as
\begin{align}
  D_r &= \partial_r - \frac{1}{r} (a\crr + b) (1 + \crr^2) \cii,  \\
  D_{\varphi} &= \partial_{\varphi} + a [(\tan 2\varphi + \sec 2\varphi) \crr^2 + \tan 2\varphi - \sec 2\varphi]
      \crr \cii + b (\tan \varphi \crr^2 - \cot \varphi) \cii,  \label{eq:D-phi-ter} 
\end{align}
in terms of $r$, $\varphi$, $\partial_r$, $\partial_{\varphi}$, $\crr$, and $\cii$.\par
%
%
Here $\crr$ and $\cii$ satisfy Eq.\ (\ref{eq:prop-1-bis}) with $k=2$, while $D_r$ and $D_{\varphi}$ fulfil Eqs.\ (\ref{eq:prop-3}), (\ref{eq:prop-2-bis}), and (\ref{eq:prop-4-bis}) with $k=2$ in the last two ones. Furthermore, it can be easily proved that
\begin{equation}
\begin{split}
  D_{\varphi}^2 &= \partial_{\varphi}^2 - 2 \left(\frac{1}{(\cos\varphi - \sin\varphi)^2} a (a - \crr^3 \cii)
      + \frac{1}{(\cos\varphi + \sin\varphi)^2} a (a - \crr \cii)\right) \\
  & \quad - \left(\frac{1}{\cos^2 \varphi} b (b - \crr^2 \cii) + \frac{1}{\sin^2 \varphi} b (b - \cii)\right)
      + 2 (a^2 + b^2 + 2ab \crr) (1 + \crr^2).   
\end{split}
\end{equation}
Hence the $BC_2$ Hamiltonian (\ref{eq:H-2}) can be extended into a $D_4$-invariant generalized Hamiltonian
\begin{equation}
\begin{split}
  \chh_2 &= - \partial_r^2 - \frac{1}{r} \partial_r - \frac{1}{r^2} [D_{\varphi}^2 - 2 (a^2 + b^2 + 2ab \crr)
        (1 + \crr^2)] + \omega^2 r^2 \\
  &= - D_r^2 - \frac{1}{r} [1 + 2 (a\crr + b) (1 + \crr^2) \cii] D_r - \frac{1}{r^2} D_{\varphi}^2 + \omega^2 r^2 
\end{split}
\end{equation}
and realizes the latter in the $D_4$ identity representation.\par
%
%
Although this result is the exact counterpart of those obtained for odd $k$ values, it has been obtained at the price of modifying the expression of $D_{\varphi}$ in terms of $\varphi$, $\crr$, and $\cii$. From Eq.\ (\ref{eq:D-phi-ter}), it is indeed obvious that the term proportional to $a$ cannot be written in a form similar to that of the corresponding one in (\ref{eq:D-phi-bis}), albeit we can re-express the term proportional to $b$ as $- b \left[\cot \varphi + \cot\left(\varphi + \frac{\pi}{2}\right) \crr^2\right] \cii$, in agreement with (\ref{eq:D-phi-bis}). Such a discrepancy is related to the above-mentioned dependence of some trigonometric identities on the parity of $k$.\par
%
%
\subsection{\boldmath Generalization to other even $k$ values}

The extension of the formalism developed for the $BC_2$ model to higher $k$ values is not so straightforward as that carried out in the previous section for odd $k$ ones. As before we start from a $D_{2k}$ group with elements $\crr$ and $\cii$ satisfying Eq.\ (\ref{eq:prop-1-bis}) and we define $D_r$ with properties (\ref{eq:prop-2-bis}) as in Eq.\ (\ref{eq:D-r-bis}). To find a generalization of $D_{\varphi}$ in (\ref{eq:D-phi-ter}), we are led by the condition that Eq.\ (\ref{eq:prop-3}) remain true. For such a purpose, it is convenient to introduce a linear combination of powers of $\crr$,
\begin{equation}
  \css = \sum_{i=0}^{(k-2)/2} \crr^{4i},
\end{equation}
which is such that
\begin{equation}
  \crr \css = \css \crr, \qquad \crr^4 \css = \css, \qquad \css^2 = \tfrac{1}{2} k \css, \qquad \cii \css = \css
  \cii, \qquad \css^{\dagger} = \css.
\end{equation}
In terms of it, we can indeed write
\begin{equation}
\begin{split}
  D_{\varphi} &= \partial_{\varphi} + a [(\tan k\varphi + \sec k\varphi) \crr^{2k-1} + (\tan k\varphi - 
      \sec k\varphi) \crr] \css \cii \\
  & \quad - b \sum_{i=0}^{k-1} \cot\left(\varphi + i \frac{\pi}{k}\right) \crr^{2i} \cii,
\end{split}
\end{equation}
which reduces to (\ref{eq:D-phi-ter}) for $k=2$ and satisfies Eqs.\ (\ref{eq:prop-3}) and (\ref{eq:prop-4-bis}).\par
%
%
On using the trigonometric identities
\begin{equation}
  \frac{1}{\left(\cos \frac{k}{2} \varphi - \sin \frac{k}{2} \varphi\right)^2} + \frac{1}{\left(\cos \frac{k}{2} 
  \varphi + \sin \frac{k}{2} \varphi\right)^2} = 2 \sec^2 k\varphi,  
\end{equation}
\begin{equation}
  \sum_{i=0}^{k-1} \csc^2 \left(\varphi + i \frac{\pi}{k}\right) = k^2 \csc^2 k\varphi,  
\end{equation}
\begin{equation}
  \sum_{i=0}^{k-1} \cot \left(\varphi + i \frac{\pi}{k}\right) = k \cot k\varphi,
\end{equation}
for even $k$, we obtain the relation 
\begin{equation}
\begin{split}
  D_{\varphi}^2 &= \partial_{\varphi}^2 - k \biggl(\frac{1}{\left(\cos \frac{k}{2} \varphi - \sin \frac{k}{2} \varphi
       \right)^2} \css a (a - \crr^{2k-1} \cii) \\
  &\quad + \frac{1}{\left(\cos \frac{k}{2} \varphi + \sin \frac{k}{2} \varphi
       \right)^2} \css a (a - \crr \cii)\biggr) \\
  &\quad - \sum_{i=0}^{k-1} \csc^2 \left(\varphi + i \frac{\pi}{k}\right) b (b - \crr^{2i} \cii) + k (a^2 + b^2 
       + 2ab \crr) \sum_{i=0}^{k-1} \crr^{2i},  
\end{split}
\end{equation}
from which it follows that the above-mentioned connection between $H_k$ and the $D_{2k}$-extended Hamiltonian $\chh_k$, defined in (\ref{eq:gen-H-k}), is also valid for any even $k$ value.\par
%
%
\section{Conclusion}

Here we have generalized the exchange operator formalism in polar coordinates, previously introduced for the CMW model \cite{khare}, to any member $H_k$, $k=1$, 2, 3,~\ldots, of the infinite family of exactly solvable and integrable quantum Hamiltonians on a plane considered in Ref.\ \cite{tremblay09a} by realizing the elements of the dihedral group $D_{2k}$ as operators on this plane. We have then constructed some differential-difference operators $D_r$ and $D_{\varphi}$, serving as building blocks for defining an infinite family of $D_{2k}$-extended and invariant Hamiltonians $\chh_k$, $k=1$, 2, 3,~\ldots. The starting Hamiltonians $H_k$ can be recovered by projecting the corresponding $\chh_k$ in the $D_{2k}$ identity representation.\par
%
%
As a final point, it is worth observing that the integrability of $H_k$, {\em i.e.}, the existence of an integral of motion $\cxx_k$ (see Eq.\ (23) of Ref.\ \cite{tremblay09a}), has an immediate counterpart for $\chh_k$ since the commuting operator $- D_{\varphi}^2$ gives back $\cxx_k - k^2 (a+b)^2$ by projection in the $D_{2k}$ identity representation. Whether the new formalism introduced here may provide a framework for proving the superintegrability conjecture of Ref.\ \cite{tremblay09a} remains an interesting open question for future investigation.\par
%
%
Another possible application of this letter might be the construction of solvable spin models associated with $H_k$ by considering the two-dimensional irreducible representations of $D_{2k}$ arising for $k > 1$.\par
%
%
\section*{Acknowledgments}

The author would like to thank A.\ V.\ Turbiner for attracting her attention to the problem addressed to in this letter and for several useful discussions. Some interesting comments of V.\ Jakubsk\'y and M.\ Znojil on Ref.\ \cite{jakubsky} are also acknowledged.\par
%
%

\end{document}